\documentclass{elsart}

\usepackage{amssymb}
\usepackage{mathrsfs}

\usepackage{epsfig}
\usepackage{amsmath}

\journal{Physica A}

\begin{document}

\begin{frontmatter}



\title{The impact of margin trading on share price evolution: A cascading failure model investigation}

\author[1,4]{Ya-Chun Gao},
\author[2]{Huai-Lin Tang},
\author[3,4,cor3]{Shi-Min Cai},
\author[5]{Jing-Jing Gao},
\author[4]{H. Eugene Stanley}
\corauth[cor3]{shimin.cai81@gmail.com (S.-M. Cai)}

\address[1]{School of Physical Electronics, University of Electronic Science and Technology of China, Cheng Du 610054, P.R. China}
\address[2]{School of Management and Economics, University of Electronic Science and Technology of China, Cheng Du 610054, P.R. China}
\address[3]{Web Sciences Center $\&$ Big Data Research Center, University of Electronic Science and Technology of China, Cheng Du 610054, P.R. China}
\address[4]{Center for Polymer Studies and Department of Physics, Boston University, Boston, MA 02215, United States of America}
\address[5]{School of information and communication engineer, University of Electronic Science and Technology of China, Cheng Du 610054, P.R. China}

\begin{abstract}
Margin trading in which investors purchase shares with money borrowed
from brokers is blamed to be a major cause of the 2015 Chinese stock market
crash. We propose a cascading failure model and examine how an increase
in margin trading increases share price vulnerability. 
The model is based on a bipartite graph of investors and shares that includes four margin trading factors, (i) initial margin $k$, (ii) minimum maintenance $r$,
(iii) volatility $v$, and (iv) diversity $s$. 
We use our model to simulate margin trading and observe how the share prices are affected by these four factors. The experimental results indicate that a stock
market can be either vulnerable or stable. A stock market is vulnerable
when an external shock can cause a cascading failure of its share
prices. It is stable when its share prices are resilient to external
shocks. Furthermore, we investigate how the cascading failure of share price is
affected by these four factors, and find that by increasing $v$ and $r$
or decreasing $k$ we increase the probability that the stock market will
experience a phase transition from stable to vulnerable. It is also found
that increasing $s$ decreases resilience and increases systematic
risk. These findings could be useful to regulators supervising margin
trading activities.

\end{abstract}

\end{frontmatter}

\section{Introduction}
\begin{figure}
\center
\includegraphics[width=\textwidth]{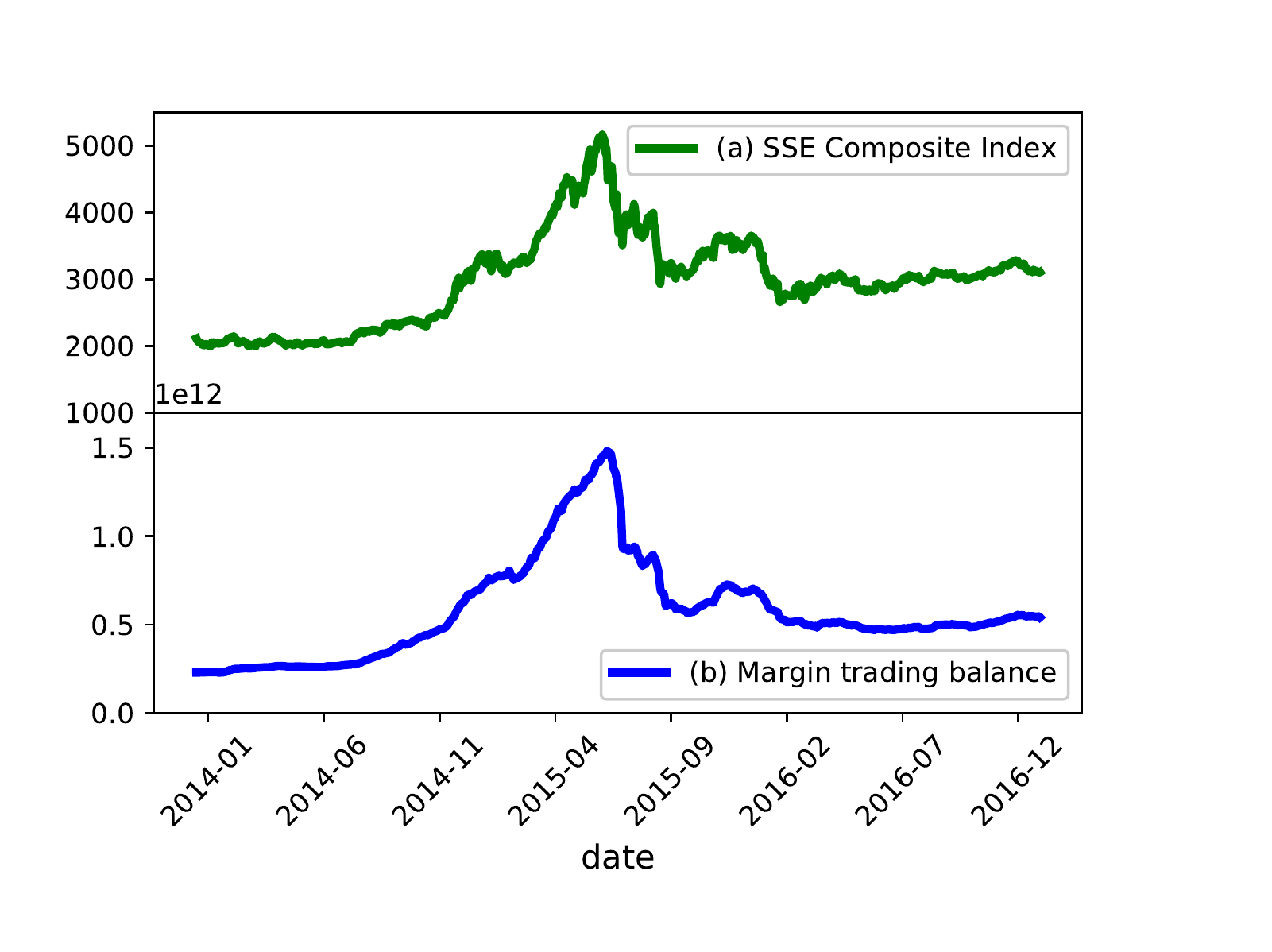}
\caption{(Color Online)(a) SSE Composite Index and (b) margin balance changed with time from the beginning of 2014 to the end of 2016.}
\label{fig:price}
\end{figure}

\hspace{0.5cm} 
During the 2014--2015 period the Chinese stock market experienced
extreme volatility and ruinous boom-bust behaviour. The important
Shanghai Stock Exchange (SSE) Composite Index rose approximately 33\% in
one month and then fell 29\% in seven trading
days. Fig.~\ref{fig:price}(a) shows that the extreme bull market began
in July 2014, that the SSE index reached a seven-year high on 12 June
2015, but that within a short period of a few weeks the same index
dropped sharply in what came to be known as the mid-2015 Chinese stock
market crash.

\hspace{0.5cm}It is speculated that this erratic Chinese market behaviour was
caused in part by a huge increase in margin trading \cite{Lu2017}.
Generally speaking, margin trading uses financial leverage.  When investors feel bullish toward an investment opportunity they borrow capital from brokers or
other resources, e.g., shadow banks, to purchase shares. To minimize
losses, brokers require investors to pay a portion of the share price as
a margin and to use the purchased shares as collateral for the
loan. There is also a requirement that there be a minimum maintenance
margin, above which the total amount of equity must be maintained in the
margin account \cite{Curley2008}.

\hspace{0.5cm}Margin trading is high risk and can yield huge profits or
total losses. Because China's securities market was immature and
approximately 90\% of its traders were retail investors, margin finance
and short-selling services were not made available prior to 2010
\cite{Sharif2014,Ma2017}. In that year the China Securities Regulatory
Commission (CSRC) conducted a pilot project and allowed shares of a few
dozen companies to be bought on margin or sold short. In September 2014,
the approved list of stocks was expanded to include more than 900
companies.
Fig.~\ref{fig:price}(b) shows that margin trading rapidly increased
and nearly doubled in the four months from September to December 2014. A
huge amount of credit was injected into the securities market and the
SSE Composite Index rapidly increased. Fig.~\ref{fig:price} shows that
the time series of the SSE Composite Index and of margin loans were
strongly correlated and fluctuated following the same trends.  During
this bull market period, share prices and margin financing activities
promoted each other, and a huge market bubble was created.

\hspace{0.5cm} Margin financing is a high-risk double-edged sword.
Although some propose that stock price behaviour indicates
\cite{Hirose2009,Saffi2011,yang2017} that margin eligibility can raise liquidity
\cite{Kahraman2014} and stabilize the market \cite{Chang2014}, others
argue that margin trading produces excess volatility and destabilizes the
market \cite{Seguin1990,Lv2017}. 
We conjecture that margin trading activities are a strong factor in the drop in stock prices during a crash and can accelerate the decline. 
Fig.~\ref{fig:price} shows that immediately following the sharp drop in share prices, margin lending also dropped sharply, which in turn accelerated the devaluation of the market index.  This behaviour was described previously
\cite{Brunnermeier2009,Thurner2012}, but the obtained results were based
on observations and regression results and did not explain the mechanism
driving the behaviour.

\hspace{0.5cm} Here we propose a cascading failure model
\cite{Dobson2007,Smart2008,Buldyrev2010,Huang2013,Hou2014} to identify the mechanism that allows margin trading to amplify the vulnerability of share price and
that caused the 2015 Chinese stock market crash.  The basic idea is that margin-covering resulting from the minimum maintenance margin requirement rapidly
decreases share price, which further triggers more margin-covering and
results in a cascading failure of share prices.

\section{Model}
\hspace{0.5cm}A bipartite graph is used to show the cascading failure
model of margin trading \cite{Souma2003,Guimera2007,Gao2013}. Nodes are
divided into two non-overlapping sets of $N=20000$ margin investors and
$M=1000$ shares. We abstract the margin trading market on this bipartite
graph and simplify the model by assigning $M$ shares the same market
capitalization and price impact factor.  
Each investor initially purchase $s$ shares on margin, with one unit of trading volume, namely, each node from the investors set is linked with $s$ nodes from the shares set. Here $s$ is a constant and $s_i=s,(i=1,2,...,N)$ for investor
$i$.  Thus the property in each margin account is the total value of $s$
shares, and all the purchased shares serve as collateral for the loan.

\hspace{0.5cm}Suppose the {\it initial margin\/} or the
{\it leverage ratio\/} is $k\in[0,1)$, i.e., the deposit of the
investor is a $k$ fraction of the whole property in the margin account,
or the investor borrows $1-k$ of the value of all purchased shares.
In the stock market, share prices constantly fluctuate. During a bull market,
investing on margin is a leverage technique that can boost profits.
When the market suddenly drops, those investing on margin may find their
margin account equity insufficient to cover their restriction of
maintenance fee. If the equity in the margin account falls below a
certain level, the investor must deposit more cash to the account to
cover the difference. This is referred to as a {\it margin call}. If the
investor does not add money quickly enough, the broker can sell the
securities without notice and liquidate the account to cover the
loan. We here denote $r$ the {\it minimum maintenance margin}. If the
ratio between the market value of the collateral and the margin loan is
less than $r$, the broker can issue a margin call.

\hspace{0.5cm}Initially, the prices of all shares are randomly assigned from a 
log-normal distribution, and the mean value of the prices is in the same magnitude of the SSE Composite Index. We designate the market index $p$ to be the
mean value of share prices, $p_{i,t}$ the price of share $i$, and
$p_t=\Sigma_{i=1}^{i=M}p_{i,t}$ the market index at time step $t$.
Although actual indices, e.g., the SSE Composite Index, are usually
capitalization-weighted, we here measure the market index using the mean
value of the share prices and assume an equal market capitalization for
all companies. At the first time step, negative factors from external
circumstances cause share prices to drop.  The negative factors could be
panic selling in response to bad news, or the prohibition of shadow
margin loans enacted by Chinese regulators on 12 June 2015 that
precipitated an increase in margin covering and reduced market
liquidity. In our model we uniformly distribute the initial price
decline $d_{i,1}$ for stock $i$ across a range of $[0,\frac{v}{100}]$,
where $v$ denotes fluctuation volatility. If $v=10$, then the price drop
is between 0 and 10\%. Thus following the initial shock the share price
at the first time step is $p_{i,t=1}=p_{i,t=0}(1-d_{i,1})$.
A number of share prices drop slightly, and a few drop sharply.  The
ensuing cascading failure is then evolved as follows:

\begin{itemize}
\item In time step $t$, the maintenance margin $r_{i,t}$ of investor $i$
  is
\begin{equation}
r_{i,t}=\dfrac{\sum\limits_{j \in \mathscr{M}_i} p_{j,t}}{\sum\limits_{j \in \mathscr{M}_i} p_{j,0}(1-k)}
\label{eq:1}
\end{equation}
where $\mathscr{M}_i$ is the set of shares held by investor $i$. The
numerator of the right-hand term is the property of each margin account,
and the denominator is the margin loan. According to the minimum
maintenance margin rule, when $r_{i,t} < r$, the broker liquidates the
account and sells all the $s$ shares belonging to $\mathscr{M}_i$ to pay
off the margin loan.

\item We then obtain the number of selling orders for each company,
  denoted $n_{i,t}^\mathrm{sell}, (i=1,2,...,M)$. The
  current market price $p_{i,t}$ is then calculated to be
\begin{equation}
p_{i,t}=p_{i,t-1}-\eta_in_{i,t}^\mathrm{sell}
\label{eq:2}
\end{equation} 
Here $\eta_i$ is the price impact factor that measures the price decline
under one unit of selling order. We set $\eta_i=\eta=5$, which means the
prices of all shares are equally impacted and decline five units under
one unit of selling order.
\end{itemize}

\hspace{0.5cm}The margin covering thus increases the number of selling orders, and the
growth of selling orders further depresses the share price, which in
turn triggers the margin call to other margin accounts.  This cascading
failure continues until the price no longer cascades, i.e., until the
range of price drop converges to a infinitesimal quantity. Here $\tau$
is the total number of cascading time steps. If margin covering does not
cause cascading failure, $\tau=1$. Otherwise, $\tau>1$.

\section{Results and Discussion}
\hspace{0.5cm}
The dynamic process is determined by \textbf{Eqs}.~(\ref{eq:1}) and (\ref{eq:2})
and the minimum maintenance margin rule.  We analyse this by applying
the mean-field method and approximating the cascading process.  Roughly
speaking, a margin investor must liquidate their position and cover
their margin when
\begin{equation}
\dfrac{\overline{p}_0(1-\overline{d})}{\overline{p}_0(1-k)} = \dfrac{1-\overline{d}}{1-k}  < r
\label{eq:3}
\end{equation}

Here $\overline{d}$ is the averaged range of share price decline at the
initial shock, and $\overline{p}_0$ is the average share price at the
initial time. Thus margin covering is affected by minimum maintenance
guarantee $r$, initial margin $k$, and $d$. According to
\textbf{Eq}.~(\ref{eq:3}), the growth of $d$ and $r$ or the reduction of $k$ will
cause a margin call to be issued \textbf{and make} the system more vulnerable to
cascading failure. On the other hand $d$ is related to $v$. Thus in our
simulation $r$, $k$, and $v$ all affect cascading failure.

\begin{figure}
\center
\includegraphics[width=\textwidth]{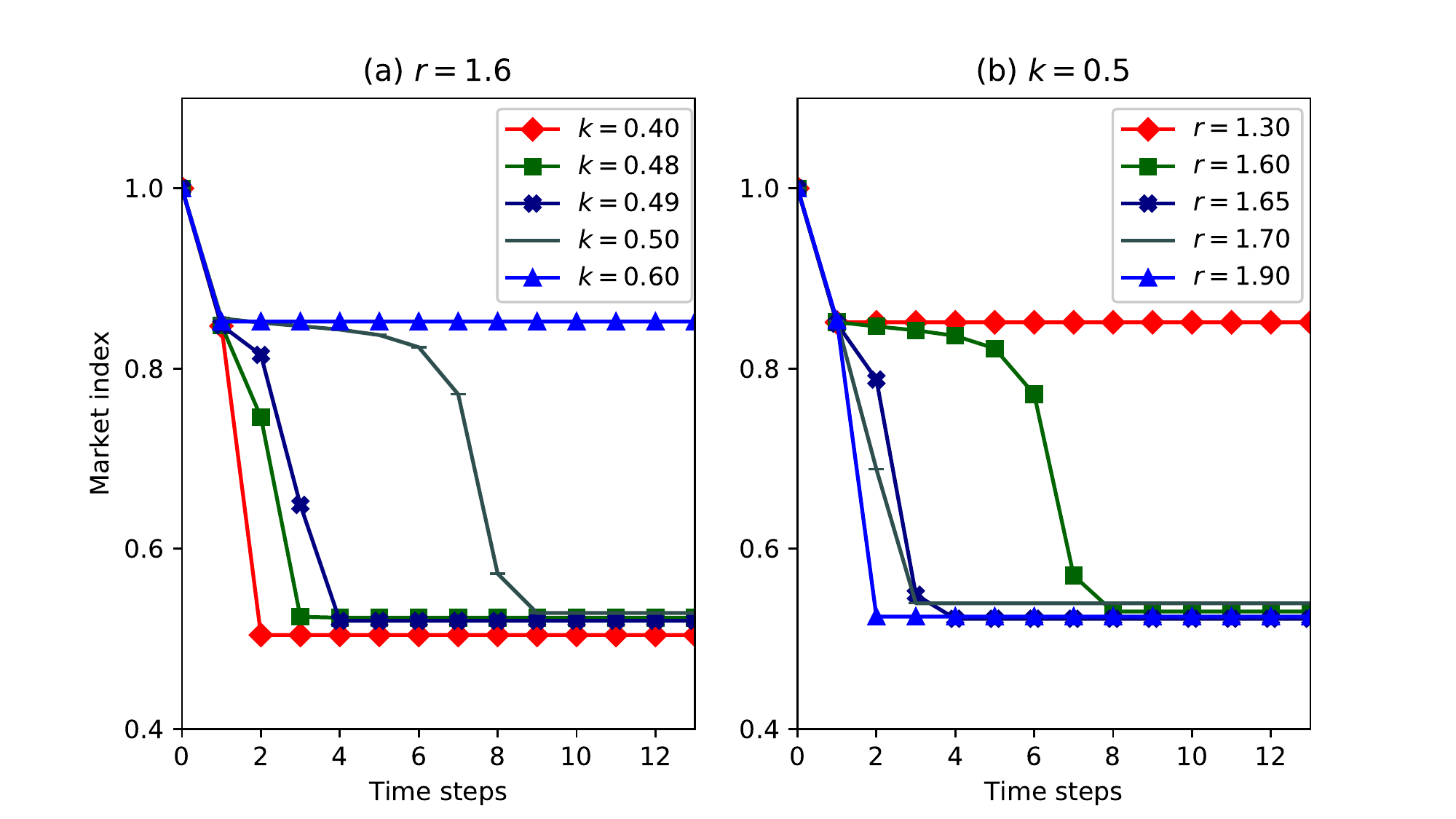}
\caption{(Color Online) Dynamical process of cascading failure with a variety of (a)$k$ and (b)$r$ values. Here, market index is the mean value of share prices. After a number of time steps, the market index reaches a steady state.
In (a), $r=1.6$, and the cascading failure evolves differently for different $k$ values, while in (b), various $r$ values are investigated at $k=0.5$. In both plots $s=20$, $v=30$. }
\label{fig:process-p}
\end{figure}

\begin{figure}
\center
\includegraphics[width=\textwidth]{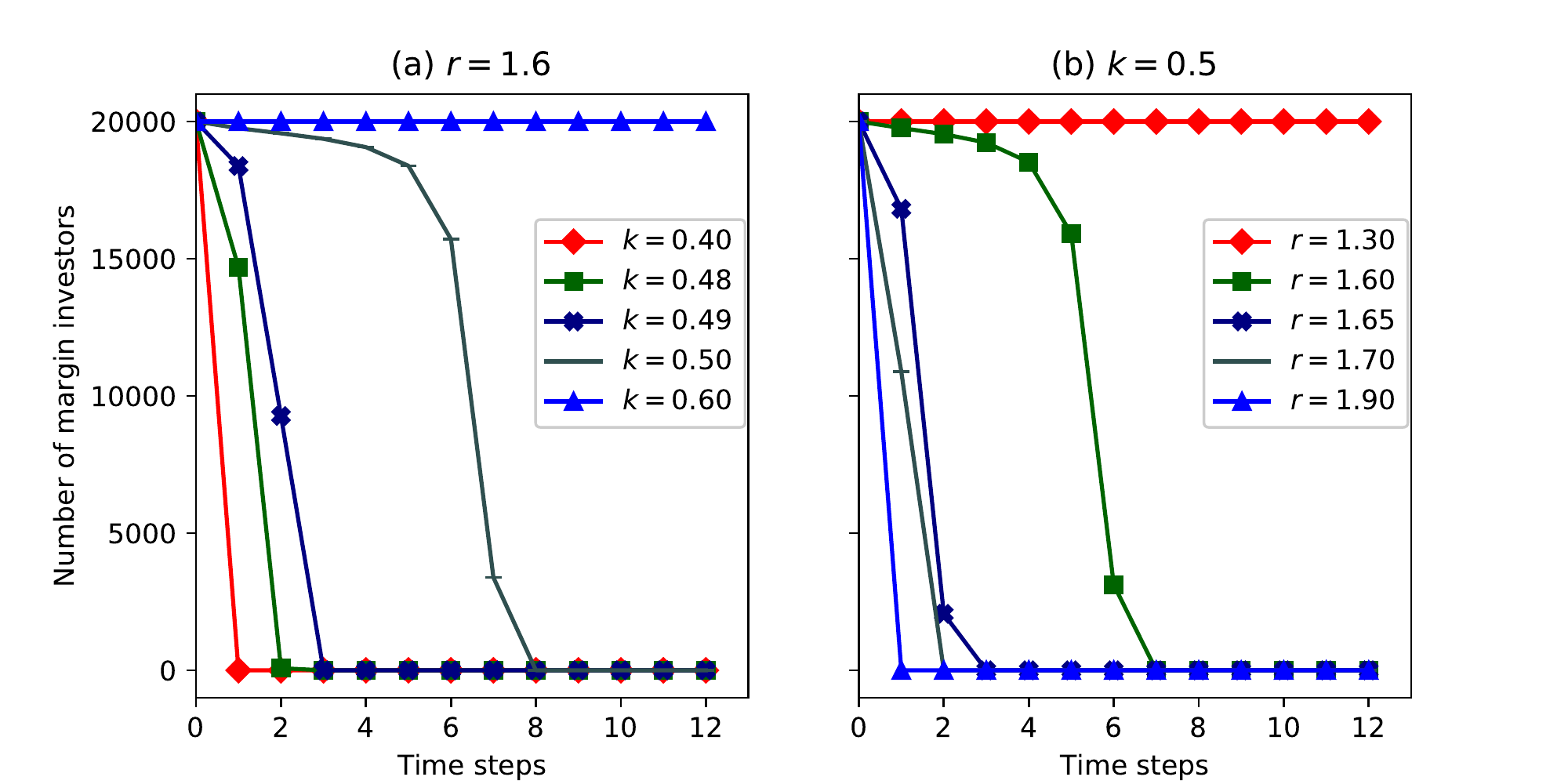}
\caption{(Color Online) Number of margin investors evolves in process of cascading failure with a variety of \textbf{(a)$k$ values when $r=1.6$ and (b)$r$ values when $k=0.5$}. Other parameters are set as $s=20$, and $v=30$.}
\label{fig:process-N}
\end{figure}

\hspace{0.5cm}Fig.~\ref{fig:process-p} plots the cascading failure process and shows
that the average price of all shares (the market index) evolves with
time. In Fig.~\ref{fig:process-p}(a), the value $r=1.6$ is set, and the simulation is carried on for a series of $k$ value, while in Fig.~\ref{fig:process-p}(b), $k=0.5$ is set, and cascading processes for different $r$ values are simulated.
In both plots $v=30$. When $k=0.6$ there is no cascading effect following the initial external shock, but when $k$ decreases to a critical value, the initial attack will cause the forced selling due to the margin call in some accounts, thus margin buying activities take places, pulling down the market index ulteriorly. 
After a small number of time steps the system converges to a steady state, all share prices remain approximately constant, and no more margin accounts are liquidated. This
is the classic cascading failure process. The market index in the steady state is $p_{\infty}$ and \textbf{the number of active margin investors} in the steady
state $N_{\infty}$. Fig.~\ref{fig:process-p}(a) shows that when
$k<0.6$ there is cascading failure. This indicates that the stock market
experiences a phase transition from a stable state to a vulnerable state
that is fragile to external shocks. Note that the minimum market index value $p_{\infty}$ seldom changes when $k=0.4\thicksim 0.5$ because in
the steady state all margin accounts have been liquidated and margin
covering reaches its maximum (see Fig.~\ref{fig:process-N}).

\begin{figure}
\center
\includegraphics[width=0.8\textwidth]{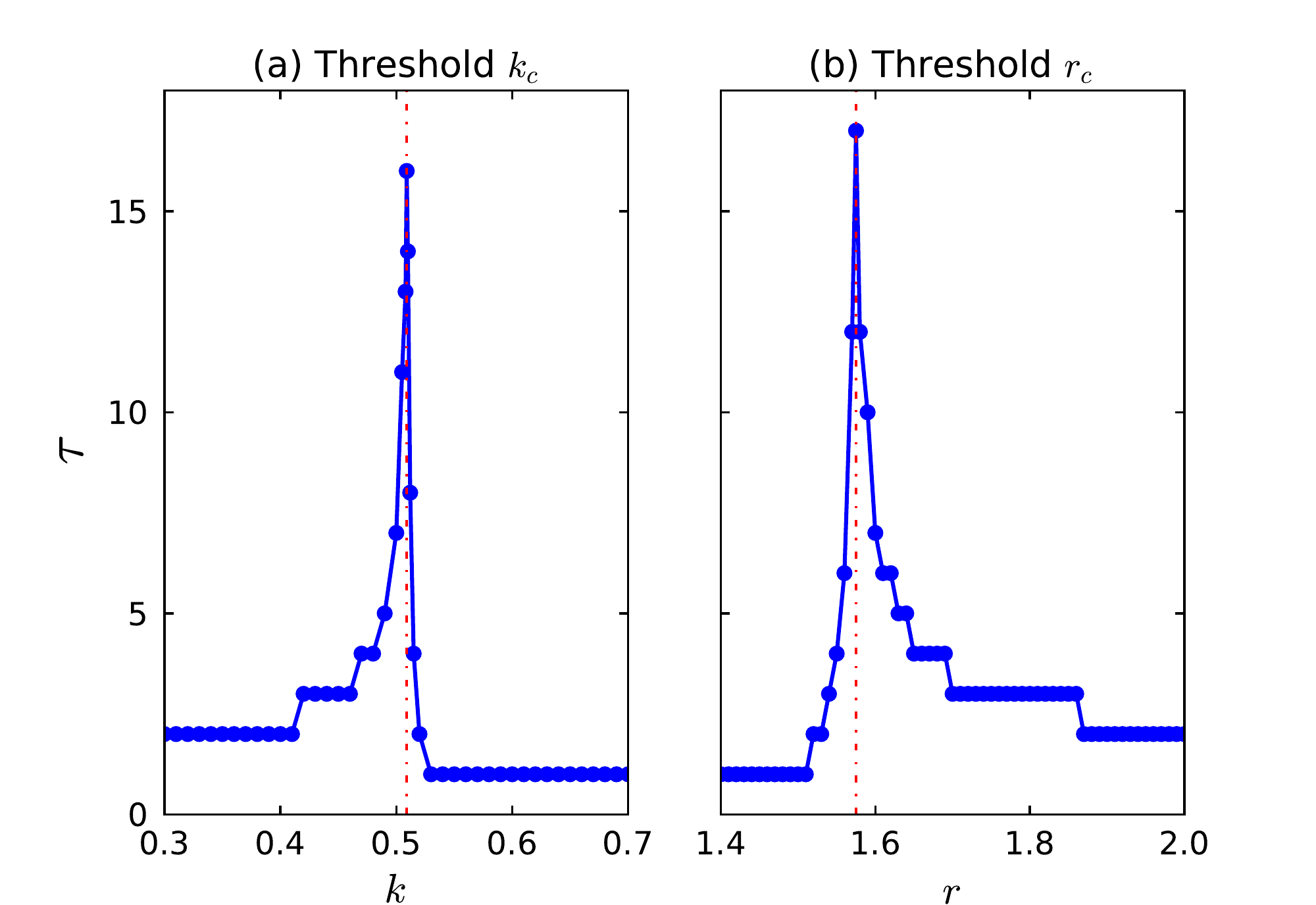}
\caption{(Color Online) Total cascading time $\tau$ as a function of \textbf{(a)$k$ values when $r=1.6$, and (b)$r$ values when $k=0.5$ , respectively}. From the curve the threshold values of the phase transition $k_c$ and $r_c$ can be recognized from the peaks. \textbf{The results are obtained by averaging over 20 simulation times.} In both plots $s=20$ and $v=30$.}
\label{fig:threshold}
\end{figure}

\begin{figure}
\center
\includegraphics[width=0.6\textwidth]{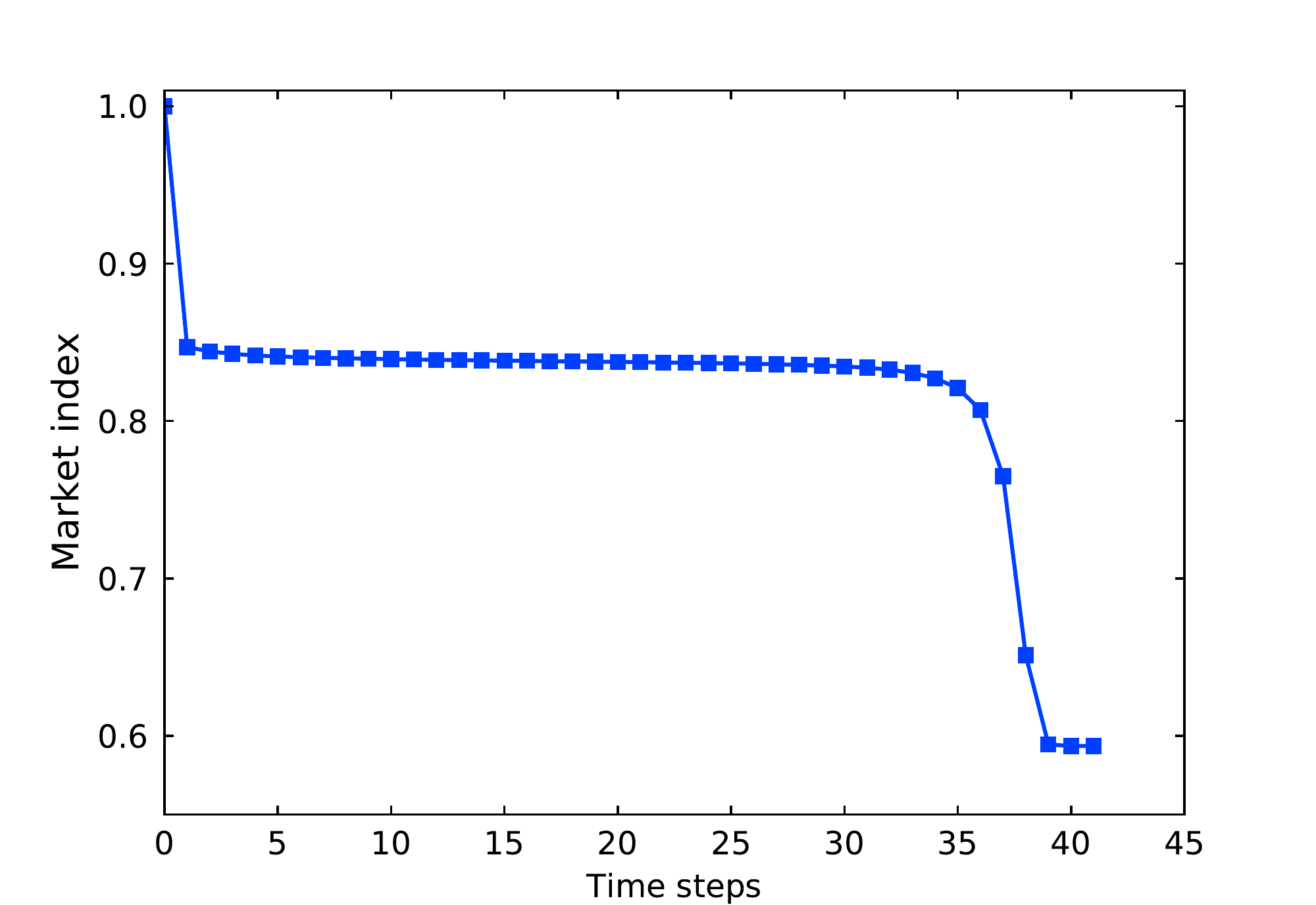}
\caption{(Color Online) The dynamic process of cascading failure at threshold of phase transition (i.e., $r=1.6$,$k=0.509$), in which a long-plateau regime is observed. Other parameters are set as $s=20$, and $v=30$.}
\label{fig:process-threshold}
\end{figure}

\hspace{0.5cm} When the $p_{\infty}$ value is unchanging, the market index drops more
rapidly when $k$ is smaller and more slowly when $k$ is large, slowing
the cascading failure process and allowing market regulators more time
to respond. Similar behaviour for various $r$ are also presented in 
Fig.~\ref{fig:process-p}(b).We then plot the total cascading time $\tau$ as a function of $k$ and $r$, respectively, as shown in Fig.~\ref{fig:threshold}. 
Fig.~\ref{fig:threshold}(a) shows that when $k$ is large there is no margin covering following an initial shock. Thus $\tau=1$. This value of $\tau$ rapidly increases as
$k$ decreases and reaches its maximum where the value of $k$ becomes
$k_c$. When $k<k_c$ the market damage is severe and after two time
steps it reaches its minimum $p_{\infty}$.  Thus $k_c$ is the critical
point of the phase transition. The dynamical process of the cascading failure at $k_c$ is presented in Fig.~\ref{fig:process-threshold}, and a long plateau stage is displayed, which is characterized by a random branching process.

\begin{figure}
\center
\includegraphics[width=\textwidth]{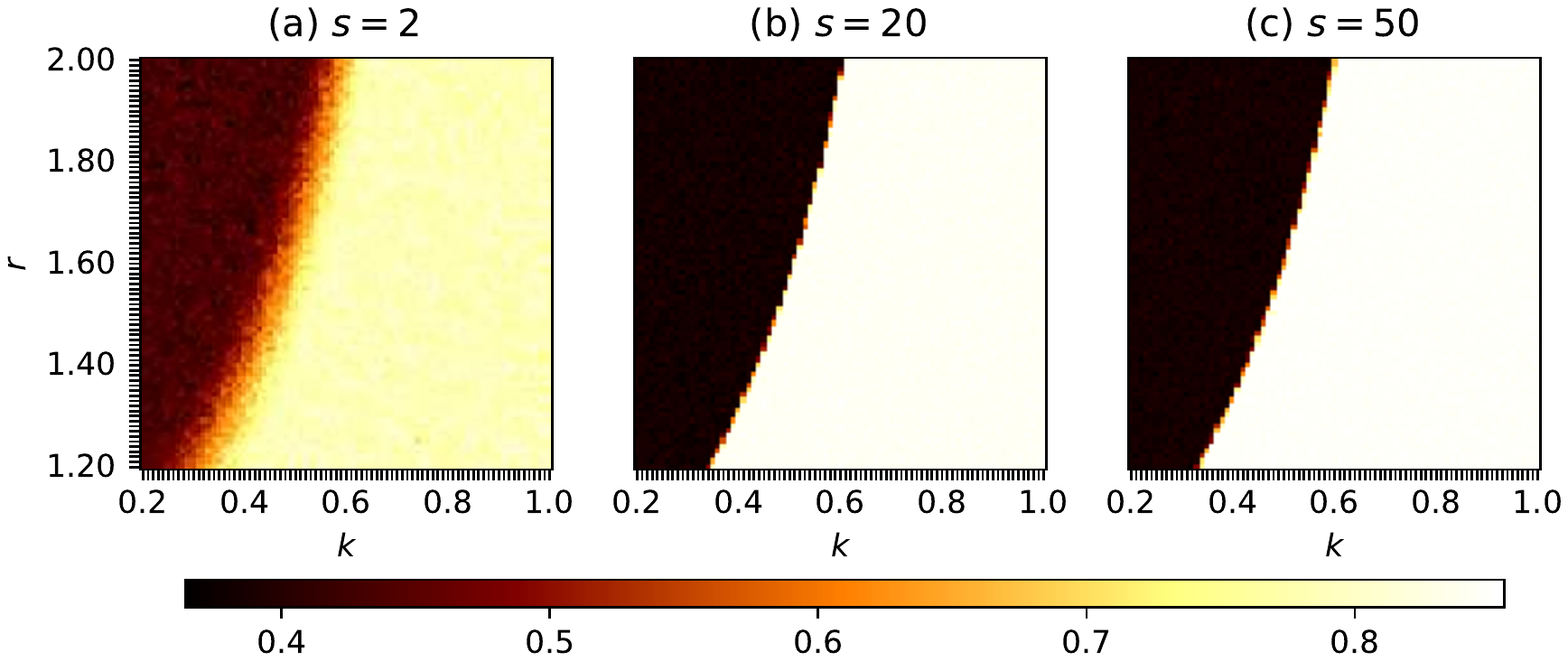}
\caption{(Color Online) Phase diagram of the \textbf{market index} as a function of maintenance guarantee $r$ and initial margin $k$. The color represents averaged price of the stock market, analogous to market index. $v=30$ for all plots.}
\label{fig:phase-rk}
\end{figure}

\hspace{0.5cm}Hence there are two attractors in the dynamic process of a bipartite
stock market: the stable and vulnerable states. 
Fig.~\ref{fig:phase-rk} plots a two-dimensional phase diagram
to verify this. The colorbar displays different $p_{\infty}$ values, and
the phase diagram shows them as a function of both $r$ and $k$, each of
which is averaged over 20 simulation times. In all three subplots with
various $s$ values, the two market states are shown. The left region
(dark color) is in the vulnerable state and is more susceptible to
cascading failure under an external shock. The right region (light
color) is in the stable state. We denote the critical values of $r$ and
$k$ to be $r_c$ and $k_c$, respectively, at which point the system
changes from a stable state to a vulnerable state, and these values are
associated with each other. Fig.~\ref{fig:phase-rk} shows that as $k$
increases $r_c$ also increases, and vice versa. When $k$ is sufficiently
large\textbf{ there is no phase transition}, and when it is small a cascading
process occurs irrespective of the $r$ value. The results are accord
with Eq.~(\ref{eq:1}).

\hspace{0.5cm}In modern portfolio theory, diversification is considered the optimal
investment strategy for lowering risk, but investor diversification may also
increases systematic risk due to the interconnectness \cite{Roukny2013,Tasca2014,Wang20171,Wang20172,Wang20173,Wang2016}. 
\textbf{In the above simulations, $s=20$.} Here,
in Fig.~\ref{fig:phase-rk}, for different $s$ the phase configuration remains the same, which means $s$ has little influence on $r_c$ and $k_c$ values.  
On the other hand, the average loss is greater when $s$ is large because the $p_{\infty}$ value in the vulnerable state is much lower when $s$ is large. The reason is if more shares are purchased on margin by each investor, the initial shock of a proportion of share prices can give rise to the contagion of more other stocks.

\begin{figure}
\center
\includegraphics[width=0.8\textwidth]{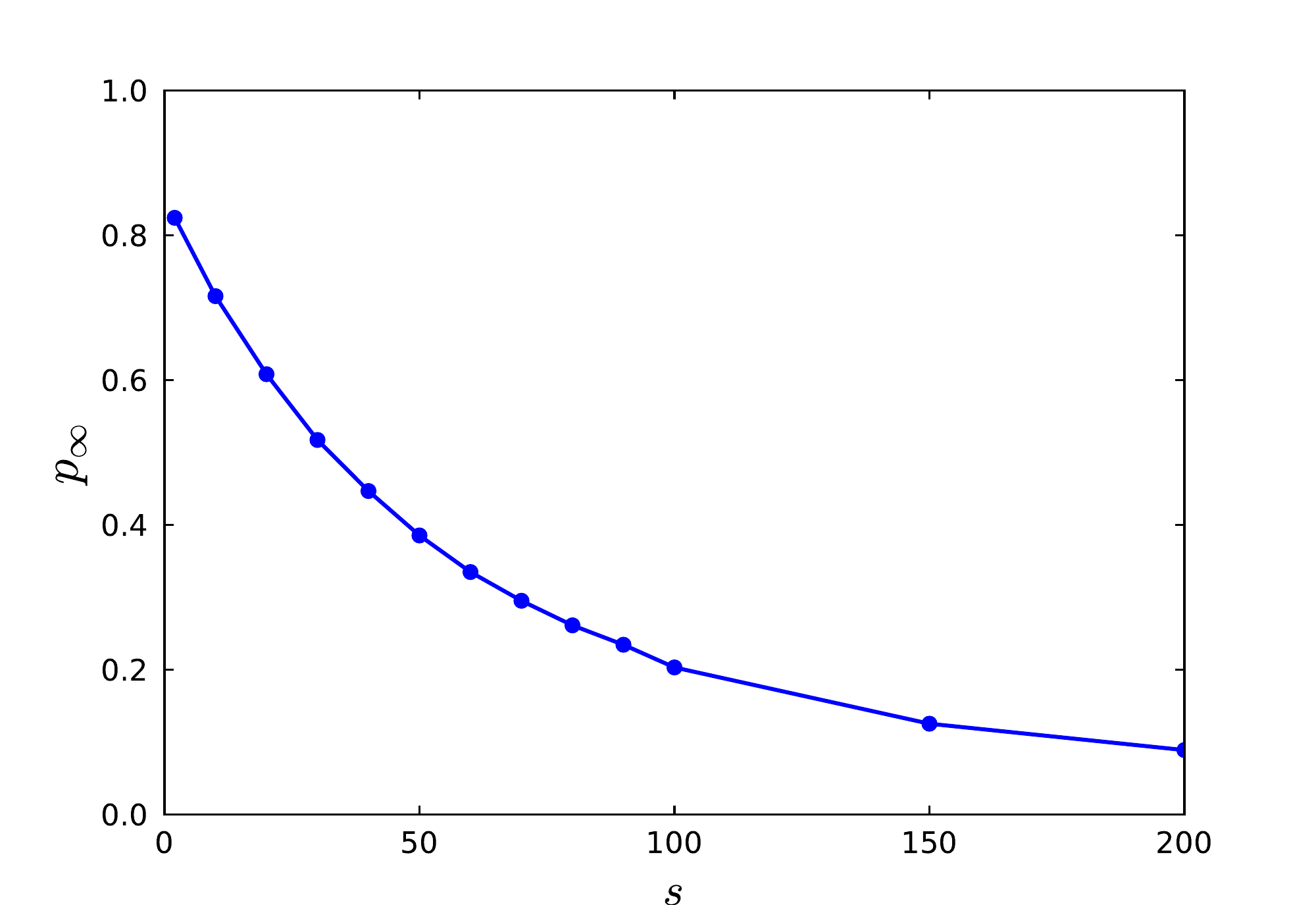}
\caption{(Color Online) $p_{\infty}$ in the steady state vs diversification parameter $s$, other parameters are set $k=0.4$,$r=1.7$,$v=30$. \textbf{The results are obtained by averaging over 20 simulation times.}}
\label{fig:phase-s}
\end{figure}

\hspace{0.5cm}To examine the impact of diversification, we calculate $p_{\infty}$
using different $s$ values (see Fig.~\ref{fig:phase-s}).  The market
bench-march index in the steady state decreases monotonically with $s$,
which indicates that diversification can lower the robustness and
resilience of the securities system and result in higher systematic
risk.

\begin{figure}
\center
\includegraphics[width=\textwidth]{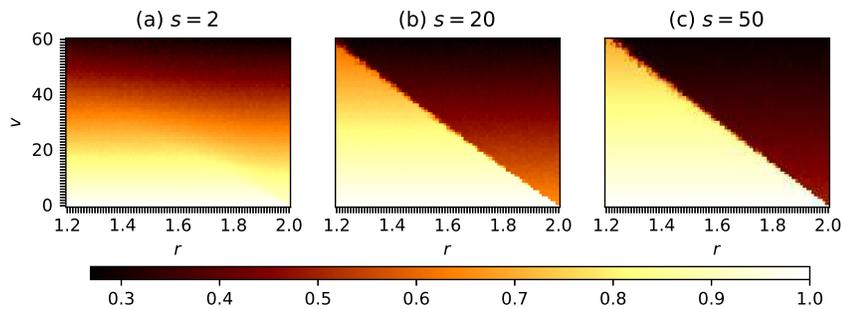}
\caption{(Color Online) Phase diagram of maintenance guarantee $r$ and volatility $v$. The color represents averaged price of the stock market, analogous to market index. \textbf{$k=0.5$for all plots, and the results are obtained by averaging over 20 simulation times.}}
\label{fig:phase-rv}
\end{figure}

\hspace{0.5cm}In recent years, the Chinese stock market has been highly volatile
because of the high proportion of active retail investors \cite{Gao2012,Gao2014, Zhao1,Zhao2,Zhao3}.  From Eq.~(\ref{eq:1}) we see that high volatility can
amplify instability and give rise to cascading
failure. Fig.~\ref{fig:phase-rv} plots the phase diagram as a function
of $r$ and $v$ to examine how volatility influence $r_c$. Note that when
$s=2$ there is no cascading failure in Fig.~\ref{fig:phase-rv}(a). This
is because risk contagion is weakened when $s$ is small, hence at some
$k$ value there is no further price decline after the initial shock, but
$r_c$ decreases when the volatility is high [see
  Figs.~\ref{fig:phase-rv}(b) and \ref{fig:phase-rv}(c)] and a failure
cascade can cause systemic failure even when the minimum maintenance $r$
is low.
Similar results are found in the $r-v$ phase diagram, and these are useful
in shaping regulatory policy in China. To stabilize the securities
market, the stock exchanges have imposed a daily price change limit of
10\% on the trading of shares of listed companies. The influence of this
price limitation remains unclear and is a topic of wide discussion
\cite{Brockman2000}. Here we proposed that this price limitation can slow
margin covering and is thus useful in stabilizing the market.

\begin{figure}
\center
\includegraphics[width=0.8\textwidth]{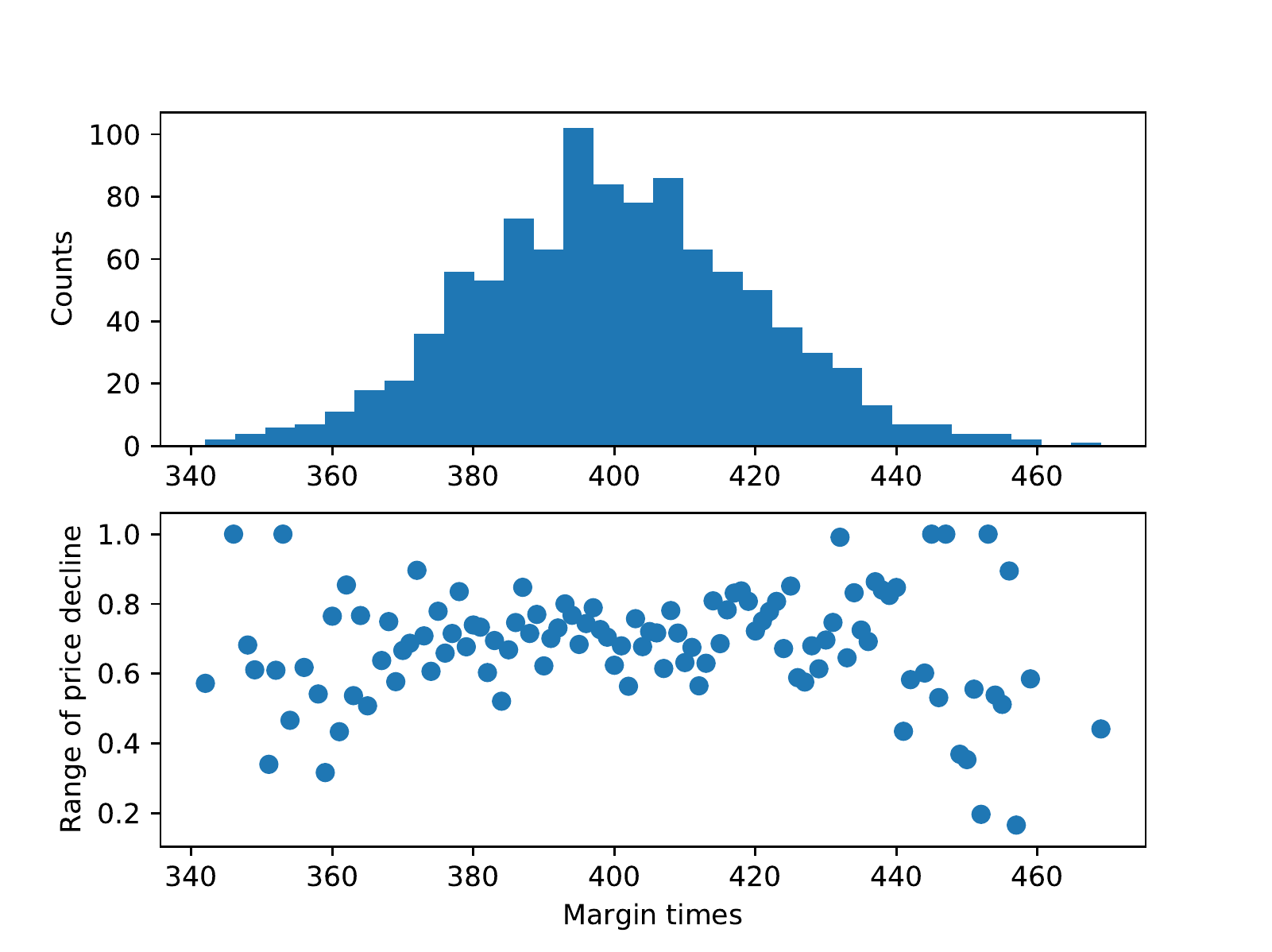}
\caption{(Color Online) The relationship between price decline and margin times. In this figure $k=0.5$, $r=1.8$, $s=20$, and $v=50$, but the results are valid for different values.}
\label{fig:stock}
\end{figure}

\hspace{0.5cm}According to the margin trading rules set by the SSE, when a stock is
bought on a margin with a proportion that is larger than 25\% of its outstanding share
capital, the securities exchange must stop the margin financing of this
stock.  This rule assumes that a share has a higher risk when the ratio
of margin buying is high. To confirm this, Fig.~\ref{fig:stock}(b)
averages the price drop of stocks \textbf{with the same margin times} in order to
link the price drop range with the margin times. The scatter plot takes
the form of a butterfly, and the results are valid for numerous $r$ and
$k$ values. Although the price drop changes very little in the middle
range of margin times, the results fluctuate up and down wildly for
stocks with small or large margin times. 
\textbf{This can be explained from the distribution of the margin times shown in Fig. \ref{fig:stock}(a), in which stocks with small or large margin times are few in number, and this affects the accuracy of the results.}

\section{Conclusion}
\hspace{0.5cm} We have examined how margin trading affected the mid-2015 stock Chinese
market crash. We use a cascading failure model---a bipartite graph of
margin investors and a set of shares---that demonstrates how margin
trading amplifies the systematic risk. 
After the initial external shock to the share price, the minimum maintenance margin
required by brokers triggers a cascading margin that depresses the share price. This
broker-induced cascading failure process can be rapid and can cause a
systemic market crash.

\hspace{0.5cm}To determine the factors influencing the cascading process, several parameters are investigated, including the initial margin $k$, the minimum
maintenance $r$, the volatility $v$, and the diversity $s$. We find two
market states, stable and vulnerable.  In the stable state, an initial
price decline affects the market but does not produce cascading failure,
and the resilient system slowly recovers.  In the vulnerable state, an
initial price decline produces cascading failure.  Both analytical and
simulation results indicate that raising $v$ and $r$ or dropping $k$
increases the probability that the state of the system will undergo a
phase transition from stable to vulnerable. We also find that diversity,
although a preferred investment strategy, can amplify systematic risk
because higher $s$ values increase stock market vulnerability.


\begin{ack}
This work is supported by the National Natural Science Foundation of China (Grant No. 71601030 and 61673086) and the Fundamental Research Funds for the Central Universities (Grant No. ZYGX2016J058). 

\end{ack}

\end{document}